\documentclass[aps,prl,preprintnumbers,a4,twocolumn,nofootinbib,showpacs ]{revtex4-1}

\newcommand{\be}{\begin{eqnarray}}
\newcommand{\ee}{\end{eqnarray}}

\def\hbar#1{\slash\hspace{-2.5mm}#1}
\def\lsim{\raise0.3ex\hbox{$\;<$\kern-0.75em\raise-1.1ex\hbox{$\sim\;$}}}
\def\gsim{\raise0.3ex\hbox{$\;>$\kern-0.75em\raise-1.1ex\hbox{$\sim\;$}}}
\def\hbar#1{\slash\hspace{-2.5mm}#1}
\usepackage{epsfig}
\usepackage{citesort}
\usepackage{graphicx}
\usepackage{dcolumn}
\usepackage{bm}
\begin{document}

\title{The geometrodynamical origin of equilibrium gravitational configurations}
\author{A. A. El Zant}
\affiliation{Centre for Theoretical Physics, Zewail City of Science and Technology, Sheikh Zayed, 12588, Giza, Egypt}%
\date{\today}
\begin{abstract}
The origin of equilibrium gravitational configurations is sought in terms of the stability of their trajectories, as described by the curvature of their Lagrangian configuration manifold. We focus on the  case of spherical systems, 
which are integrable in the collisionless  (mean field) limit despite the apparent persistence of 
local  instability of trajectories even as $N \rightarrow \infty$.
It is shown that when  the singularity in the potential is removed, a null scalar curvature is associated with
an effective, averaged, equation of state describing dynamically relaxed equilibria with marginally stable 
trajectories. The associated configurations  are  quite similar to those of observed elliptical galaxies and  simulated cosmological halos. This is the case because a system starting far from equilibrium
finally  settles in a state which is integrable when unperturbed, but where it can most
efficiently wash out perturbations. We explicitly test this interpretation by means of direct simulations.
\end{abstract}

\maketitle

\section{Introduction}

  In a contemporary context, the Newtonian dynamics of a large number of self-gravitating particles describes 
the dynamics of putative dark matter, and to first order the formation and 
evolution of galaxies and clusters. Galaxies
that are thought to have formed through largely dissipationless processes 
show remarkable regularity in their density profiles, and dark matter halos identified in 
cosmological simulations have similar (and  nearly) universal profiles~\cite{NFW, MerNav}. 
Yet there is no satisfactory theory predicting these preferred choices of dynamical 
equilibria. 

Steady states are characterized by the absence of coherent  modes, which
decay due to differences  between motions on neighboring 
trajectories. This  effect is present in all nonlinear systems (phase mixing [3]),
but is  most efficient if nearby trajectories rapidly diverge and spread in phase space.
However, attempts at relating these trajectory stability properties to the 
choice of $N$-body equilibrium --- or extracting almost any interesting information at all from them ---
are frustrated by a puzzling phenomenon.
As $N \rightarrow \infty$,  particles move 
in the mean field potential, and steady state systems with sufficient symmetry become 
near integrable --- e.g., spherical systems are completely integrable when in equilibrium, 
and phase space distances between their trajectories thus generally diverge linearly 
in time~\cite{BT}. In contrast, it has long been
realized that direct integrations of the equations of motion invariably reveal exponential divergences on a timescale that does not
increase with $N$~[4, 5]. 

   A clue concerning the resolution of this
paradox comes from studies of softened systems. For when
the singularity in the potential is removed (e.g. $\phi \sim1/r \rightarrow  \phi \sim 1/\sqrt{r^2+  {\rm const}}$), the
divergence timescale does increase with $N$ [6, 7].
The purpose of the present study is to show that, in this case, 
physically interesting information, pertaining to the 
choice of equilibrium, 
can be extracted from the stability
properties of trajectories.

\section{Equilibrium and pressure support}

  The divergence of nearby trajectories can also be deduced 
from  geometric representations of  motion and stability in 
gravitational systems 
(e.g., \cite{GS, Kandrup90, El-Zant97, CEPett}). 
Though the geometric description is not unique~\cite{Casetti:2000gd}, one well known
formulation has the advantage of involving a positive definite metric 
defined directly on the Lagrangian configuration manifold ---  a  curved subspace  
of the  $3N$-dimensional space spanned by Cartesian particle coordinates, the phase space being its 
cotangent bundle~\cite{Arnold}. This conformally flat metric is expressed in 
terms of the Cartesian  coordinates and the system kinetic energy,  $W =  E - V$, by
$
ds^{2}= W \sum_{3N} \left(dx^{\alpha} \right)^{2} \label{eq:L2}
$  (e.g., \cite{Arnold, Lanczos}).
   The divergence of  trajectories is then determined by the two dimensional sectional
curvatures $k_{\bf u,n}$, where ${\bf u}$ is the system velocity vector and ${\bf n}$ 
define directions normal to it. When the  $k_{\bf u,n}$ are negative, trajectories are exponentially unstable~\cite{Anosov}. 
Isotropic $N$-body systems with singular two-body potentials are predicted to tend toward this 
state as $N$ is increased and direct two-body collisions are ignored~\cite{GS, Kandrup90}; and
trends similar to those obtained from direct calculations 
are recovered when the full gravitational field  is taken into 
account~\cite{Kandrup90}. 

We will be interested in how $N$-body systems damp out fluctuations, we therefore  
seek an averaged measure of  their response to random perturbations.
Integrating over directions ${\bf n}$, 
we obtain the  mean (or Ricci) curvature $r_{\bf u}=\sum_{\mu=1}^{3N-1} k_{ {\bf n_{\mu},u} }(s)$. 
Furthermore, we will focus on isotropic systems. In this case, a further average
over the velocity directions ${\bf u}$ can be made, giving
the Ricci scalar; which, for $N \gg 1$, can be expressed as       
\begin{equation}
R=\sum_{\bf u,n} k_{\bf u,n} = - 3 N~\frac{\nabla^{2}W}{W^{2}} 
-  9 N^2~\frac{\parallel \nabla W \parallel^{2}}{4 W^{3}},
\label{scalar}
\end{equation}
where $\nabla W$ and $\nabla^{2} W$ are the gradient and Laplacian,  
taken with respect to coordinates $q_i = \sqrt{m_p x_i}$, and
$m_p$ are the particle masses (which we will assume to be equal).
For  sufficiently large-$N$ spherical systems with isotropic velocities 
$r_u \rightarrow R/N$ ~\cite{ElZant:1998uf}. 

A correspondingly averaged equation, describing 
stability in the time domain, is given  by~(e.g., \cite{CEPett})
\begin{equation}
\frac{d^2 Y}{dt^2} + Q(t) Y = 0.
\label{eq:stab}
\end{equation}
Here $Y$  measures the mean divergence of trajectories and 
\begin{equation}
Q = \frac{2 W^2}{9N^2}~R - \frac{1}{4} \left(\frac{\dot{W}}{W}\right)^2 
+ \frac{1}{2} \frac{d}{dt} \left(\frac{\dot{W}}{W}\right).
\label{eq:flucs}
\end{equation}
Near dynamical equilibrium, $\dot{W} \rightarrow 0$ and the dominant term on
the right hand side will be the curvature term. 
For singular Newtonian potentials, and when direct collisions are ignored, 
Eq.~(\ref{scalar}) predicts negative $R$ (since the first term vanishes), and 
exponential instability is present even in the case of  spherical equilibria. 
Ommitng direct collisions however implies an incomplete manifold; 
it is not clear if straightforward
application of the geometric approach remains appropriate in this case~\cite{ABMARDS}. 

Formally, 
as $N \rightarrow \infty$, the first term in Eq.~(\ref{scalar})
would contribute through  a collection of delta functions; but if the singularity in the potential is 
softened, this term represent a  smoothed out density field integrated over the particle distribution
(with the softening length acting as smoothing parameter), and, in dynamical equilibrium, the two terms on the right hand side of Eq.~(\ref{scalar})
become comparable. 
Indeed, by multiplying the terms  by $W^2/N^2$, taking the inverse, 
and the square root, these terms can be seen to be linked to 
the natural timescale of the system (the dynamical, or crossing, time) ---  
the first through the relation $\tau_D \sim  1/ (G \bar{\rho})^{1/2}$ (where the average density is related to the Laplacian via the Poisson equation), and the second through $\tau_D \sim (W / m_p)^{1/2}  \bar{a}^{-1}$, where $\bar{a}$
is the average acceleration affecting a  particle.  

In fact, the  structure of Eq.~(\ref{scalar})  reflects 
a pressure-gravity balance. To see this, assume $R=0$. Then
\begin{equation}
4 \pi G~W~\sum_{1}^N \rho_i = \frac{3 }{4}~ N ~ m_p  \sum_{1}^N a_i^2,
\end{equation}
where $\rho_i$ and $a_i$ refer to the density and acceleration at the position of particle $i$.   
This equates a density multiplied by the velocities squared (i.e, a pressure
term) to a gravitational binding term. In the continuum approximation, and for a spherical
system, this becomes
\begin{equation}
4 \pi W \int \rho^2 r^2 dr = \frac{3}{4}  M G \int \frac{m^2}{r^2} \rho dr, 
\end{equation} 
where $\rho = \rho (r)$ is the radial density, $m$  is the mass enclosed within radius $r$,
$M$ is the total  mass, and the integrals are evaluated over the volume of the 
configuration.   In dynamical equilibrium 
one can use the Jeans equation for an isotropic system (e.g., ~\cite{BT}) to get
\begin{equation}
4 \pi W \int \rho^2 r^2 dr = -  \frac{3}{4}  M \int m~d (\rho \sigma^2), 
\label{curcrude}
\end{equation} 
where $\sigma^2 = \bar{v^2} (r)$ is the one-dimensional velocity dispersion.
There is one easily identifiable solution of Eq.~(\ref{curcrude}); it corresponds to  $\rho \sim 1/r^2$ and $\sigma = {\rm constant}$. This is a  singular isothermal sphere; though its gravitational 
force, and both terms in Eq.~(\ref{curcrude}), diverge as $r \rightarrow 0$,
and the mass diverges as $r \rightarrow \infty$, there is significance to this solution, as isothermal spheres enclosed by a boundary are
entropy maxima of softened gravitational systems, which
do tend toward these maxima, provided certain conditions are met~\cite{DLBW, El-Zant98}.

For well behaved systems, $\rho \sigma^2=0$ as $r \rightarrow 0$ and $r \rightarrow  \infty$. Integrating by parts, noting this, and  
that the kinetic energy is related to the average  velocity dispersion by $W = \frac{3}{2} M \langle \sigma^2 \rangle_r$, and $dm =4  \pi r^2  \rho dr$, one gets
\begin{equation}
\langle{\sigma^2} \rangle_r ~ \int \rho~dm = \frac{1}{2} \int \rho \sigma^2~dm.
\label{fund}
\end{equation}
If well defined pressure and temperature functions can be assigned (i.e.,  assuming  local thermodynamic equilibrium), 
then $P = \rho \sigma^2$ and $W= \frac{3}{2} N k~ \langle T \rangle_r$, and 
\begin{equation}
\int P~dm = 2~\frac{k~\langle T \rangle_r }{m_p} \int \rho~dm. 
\end{equation}
Save for the integrations over the mass distribution, this form is that of an ideal gas equation of state.  
Nevertheless, because of the factor  $2$, there are no associated isothermal states; in contrast, 
solutions require that the density and velocity dispersion distributions correlate.

\section{Approach to EQUILIBRIUM: an interpretation}

The results obtained thus far suggest that {dynamically relaxed equilibrium systems should 
have small or vanishing $R$}; but how do they reach such configurations?  

The density and potential of a system started far from
equilibrium will  fluctuate until  collective 
motions are damped;  this will take place most 
efficiently when it finds a configuration where its response is least coherent; 
that is,  when the divergence of neighboring trajectories are maximal, and coherent modes 
are damped through phase space mixing of trajectories as the 
system responds to density and potential fluctuations. 
At the same time, it is required that in the absence of fluctuations the trajectories of
a spherical system should not be unstable, so as to recover an integrable system in the inﬁnite-$N$ limit. 
This implies that $R$ cannot be negative at equilibrium.

Eq.~(\ref{eq:stab}) suggests that configurations that satisfy these requirements have
$R \rightarrow 0$ at equilibrium: a system starting far from equilibrium should
then end up in a configuration satisfying this
condition. To show this quantitatively, we have used a formulation, developed by Pettini and collaborators, which
assumes that, for positive $Q$,  unstable solutions of Eq.~(\ref{eq:stab}) can be 
estimated by 
 dividing  $Q$ into a mean term $k_0$ and rms fluctuations $\sigma_k$  with characteristic timescale $\tau$; this gives 
rise to an effective  Lyapunov exponent  (e.g., Eqs. 79 of~\cite{Casetti:2000gd})
\begin{equation}
\lambda = \frac{1}{2} \left(\Lambda - \frac{4 k_0}{3 \Lambda}\right),
\end{equation}
with
$\Lambda^3 = \sigma^2_k \tau + \sqrt{\left(\frac{4 k_0}{3}\right)^3 + \sigma^4_k \tau^2}$.

In general, phase space averages are not well defined for gravitational systems (which 
have a non-compact phase space). Nevertheless, every  exact  ($N \rightarrow \infty$)
collisionless equilibrium has a characteristic $R$  and associated  
$k_0 =  \frac{2 W^2  R}{9~ N^2}$.
It is therefore possible to quantify the effect of fluctuations
of a certain magnitude $\sigma^2_k = \langle Q^2 \rangle - \langle Q \rangle^2$ about a given equilibrium, 
and to compare the magnitude  of this effect for different collisionless configurations.  

The natural timescale for fluctuations in a gravitational system is the dynamical (or crossing) time, so
$\tau = \tau_D$.  And since $Q$ comes in units of inverse time squared, we can write
$k_0 = a/\tau_D^2$ and $\sigma = b/\tau_D^2$, and express the exponentiation 
time associated with trajectory divergence in terms of the natural timescale as 
\begin{equation}
\frac{\tau_e}{\tau_D} =   \frac{6 \left(b^2 + \sqrt{\left( \frac{4}{3} a \right)^3 + b^4}\right)^{1/3}}
{3 \left(b^2 + \sqrt{\left( \frac{4}{3} a \right)^3 + b^4}\right)^{2/3} - 4 a}.
\label{petstab}
\end{equation}
This is proportional to  $a b^{-2}$ for $a \gg b$ and 
tends to  $2^{2/3} b^{-2/3}$ as $b \gg a$.
For  a given fluctuation level $b$, the decoherence of trajectories due to local divergence will 
be more efficient,  i.e. its exponential time will be smaller, for smaller $a$,
and therefore $k_0$ and average $R$.

\section{Testing the interpretation}

   We now examine the above interpretation 
by means of direct $N$-body simulations. Since we are interested
in estimating the densities and accelerations in the continuum limit, we use a technique~\cite{HERQOS},
which expands  the density and potential in smooth functional series; 
and since we are interested in strictly spherical configurations,
we  only make use of the radial expansions in this scheme (which 
is carried up to order $30$). Systems are sampled using a $100 000$ particles, 
started from homogeneous spatial initial 
conditions inside a unit sphere of unit mass. This configuration has  
a natural timescale $\tau_D  = (G \rho (0))^{-1/2} = \sqrt{4 \pi / 3}$, which
will be our time unit.

We have run configurations starting with  isotropic velocities that are either constant 
or following various decreasing or increasing functions of radius. 
The results shown here correspond to the latter case, for the following reason.
 Collisionless evolution implies that 
the  'pseudo-phase space density' $\rho_p = \rho/\sigma^3$  generally 
decreases along particle trajectories~\cite{BT}; and one of our aims  is to explain the dynamical origin of     
cosmological halos, which have  (a centrally divergent) $\rho_p \sim r^{-1.875}$~\cite{TN}. 
From among  various functional forms tried, results will be shown for
initial $\sigma \sim r$, noting that the trends reported seem generic.

We vary the initial virial ratio ${\rm Vir} = 2 W/ |V|$ from $1$, corresponding to near 
equilibrium, to $0.125$, corresponding to a system started far from equilibrium.
Given the sampling errors for finite $N$, $R$
will never tend to zero exactly in practice; and the absolute values of the 
densities and accelerations (and local dynamical times) strongly 
depend on the central concentrations of the final configurations. 
A relative measure is thus required for comparing these. We use the normalized 
quantity 
\begin{equation}
R_n = 
\left(4 \pi G W \sum_1^N \rho_i\right) \left(\frac{3 }{4}~ N ~ m_p  \sum_1^N a_i^2 \right)^{-1}
 - 1,  
\label{Rn}
\end{equation}
which  measures the departure from equality in Eq.~(4).

\begin{figure}[t]
\begin{center}
\epsfig{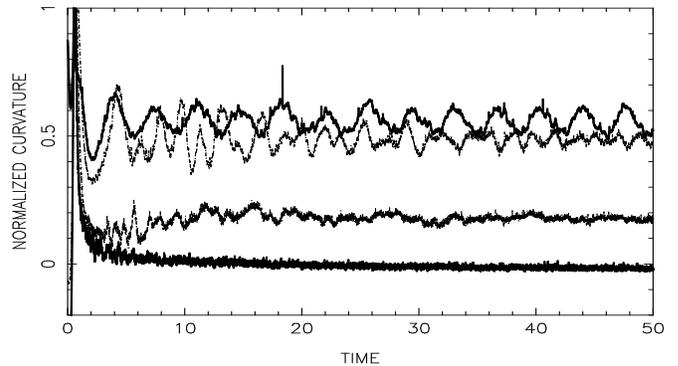}
\end{center}
\vspace{-0.6cm} %
\caption{Time evolution of normalized curvature $R_n$ (Eq.~\ref{Rn}), measured 
in  initial dynamical times, for systems starting with (from top to bottom) 
virial ratios, 1, 0.5, 0.25 and 0.125} 
\label{curvatures}
\end{figure}
\begin{figure}[t]
\begin{center}
\epsfig{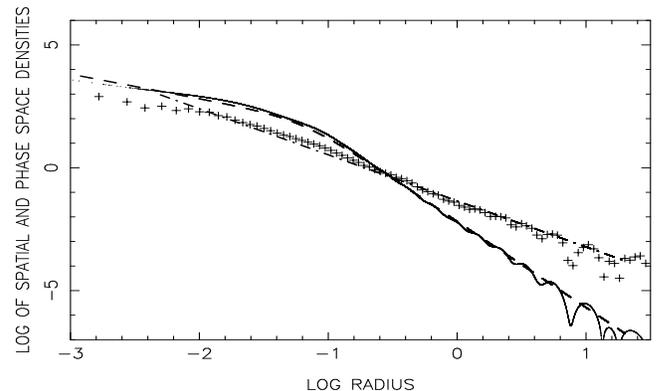}
\end{center}
\vspace{-0.6cm} %
\caption{Equilibrium density (solid line) and phase space density
(crosses) of system starting with virial ratio 0.125. Dashed lines
are eye fits with$\rho \sim 1/r (1+r^{2.5})$ and $\rho_p \sim r^{−1.875}$} 
\label{densfit}
\end{figure}

  The results are in Fig.~1, which shows that the  equilibrium $R_n$ is indeed always of order unity or 
less, and that the further from equilibrium the initial state, the closer the evolved system 
is to a configuration that minimizes $R$. The initial $R_n$ is zero when ${\rm Vir} =0.5$;
systems  with initial ${\rm Vir}< 0.5$ start 
with $R < 0$ and then explore state space until they achieve configurations where (according to Eq. ~\ref{petstab})
oscillations are efficiently damped.

Also, as shown in Fig.~2, the equilibrium density and  phase space density profiles 
of this  system are remarkably similar to those of cosmological halos (and shallow cusp elliptical galaxies~\cite{MerNav}).
There are some differences. Our outer density profile is steeper than the $1/r^3$ form of
cosmological halos; however finite configurations cannot have  $1/r^3$ outer  profiles. (Cosmological halos are not isolated equilibrium structures as their outer regions are subject to mass infall.)
The functional form of our empirical spatial density fit is also slightly different from
standard 'NFW'  (where $\rho_{\rm NFW} \sim \frac{1}{r (1 + r)^2)}$), and there seems to
be some flattening toward the center of the $1/r$ cusp. Yet, similar departures from ideal NFW often also apply to 
cosmological halos [22].

   Finally, we explicitly verify  that
configurations with smaller $R_n$  do
efficiently damp out fluctuations. This
is already  apparent in Fig.~1, where ﬂuctuations are quickly damped for 
such systems. (This may be intuitively expected, for these systems lack
homogeneous harmonic cores with nearly constant orbital frequencies and coherent response.) 
Further verification
is obtained by perturbing the final states and measuring
variations in the density distributions as  
systems, so perturbed, are evolved. The results 
of one such experiment, where the spatial spatial coordinates were decreased by $10 \%$,
are shown in Fig.~3. There, the system that was perturbed from a state with smaller 
$R_n$  is seen to quickly settle back to an equilibrium that is closer to the unperturbed one, 
and to gradually close in further on it (as the outer regions, with large crossing times, evolve).

\begin{figure}[t]
\begin{center}
\epsfig{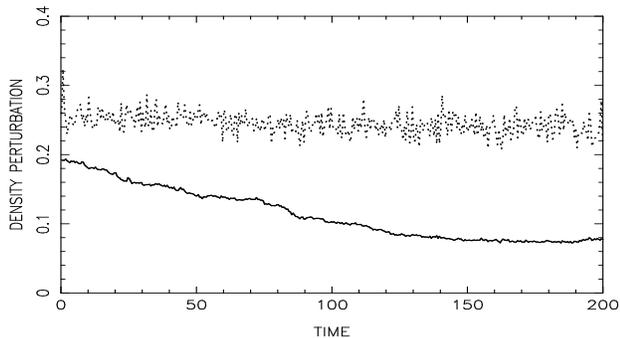}
\end{center}
\vspace{-.6cm} %
\caption{Evolution of density perturbations, defined as $\sum_{i=1}^{N} (\rho_i (t = 0)  -  \rho_i (t))^2 /\sum_{i=1}^{N} \rho_i^2 (t = 0)$, when coordinates of final states of systems  starting from  virial ratio 0.125 (solid line)  and  1 (dotted line) are perturbed by $10 \%$} 
\label{perturb}
\end{figure}


\section{Conclusion}
   
Contradictions between  predictions  pertaining to the dynamical stability
of large-$N$ body system trajectories  and those 
implied by the collisionless limit disappear when the singularity in the potential is
removed. What transpires instead is a description of gravitational equilibria.
 We propose an interpretation of the approach to equilibrium that is effectively a  rationalization of the classic 
idea of 'violent relaxation' [23]:  systems starting far
from equilibrium reach such a state once a configuration where 
collective oscillations are efficiently damped is reached, and such 
configurations  correspond to states whose dynamical trajectories are marginally stable. 
Imposing this condition leads to relations describing relaxed dissipationaless equilibrium configurations that are 
remarkably similar to those observed and modeled.  
These systems do not 'ring' and are more robust.

\end{document}